# FILAMENT ERUPTION ON 2010 OCTOBER 21 FROM THREE VIEWPOINTS

Boris Filippov

Pushkov Institute of Terrestrial Magnetism, Ionosphere and Radio Wave Propagation of the Russian Academy of Sciences (IZMIRAN), Troitsk, Moscow, 142190, Russia; bfilip@izmiran.ru

## ABSTRACT

A filament eruption on 2010 October 21 observed from three different viewpoints by the *Solar Terrestrial Relations Observatory* (*STEREO*) and the *Solar Dynamic Observatory* (*SDO*) is analyzed with invoking also data from the *Solar and Heliospheric Observatory* (*SOHO*) and the Kanzelhoehe Solar Observatory. The position of the filament just before the eruption at the central meridian not far from the center of the solar disk was favorable for photospheric magnetic field measurements in the area below the filament. Because of this, we were able to calculate with high precision the distribution of the coronal potential magnetic field near the filament. We found that the filament began to erupt when it approached the height in the corona where the magnetic field decay index was greater than one. We determined also that during the initial stage of the eruption the filament moved along the magnetic neutral surface.

## 1. INTRODUCTION

Solar filament eruptions are most conspicuous phenomena in the low and middle corona that precede the observation of coronal mass ejections (CMEs) in the upper corona. Like CMEs, they show the ascending motion of coronal plasma, although different spectral ranges and instruments of different designs are used to reveal the motion in these two coronal domains. However, there is no one-to-one correspondence between eruptive prominences and CMEs. Statistical studies show different correlations between CMEs and prominence eruptions ranging from 10-30% (Yang and Wang 2002) to 82% (Gopalswamy et al. 2003) and 92% (Hori & Culhane 2002). These discrepancies are likely due to the choice of data used in the correlation analysis and strict definition of the terms. There are failed filament eruptions when filament rising motions stop at some higher altitude (Vrsnak et al. 1990; Filippov & Koutchmy 2002; Török & Kliem 2005; Alexander et al. 2006). Typically, such events are not associated with a CME. In successful filament eruptions, the filament remnants usually constitute a bright core of a CME. On the other hand, a pre-eruptive CME magnetic structure may not contain enough cool and dense plasma to be visible as a prominence or a filament. At present, observations of filament eruptions on the disk are more difficult than observations of CMEs in the outer corona, so some eruptive filaments can elude from observers.

The rate of energy supply for acceleration of the CME material with a mass of ~ $10^{16}$ g to the speed of ~ 1000 km s$^{-1}$ within about half an hour is about $10^{29}$ erg s$^{-1}$. It is large compared to the rate of magnetic energy injection into the corona through the photosphere in an active region, which is ~ $5 \cdot 10^{27}$ erg s$^{-1}$ (Régnier & Canfield 2006). So, the energy should be accumulated and stored in the corona over a long time period before the eruption. Coronal electric currents are generally considered as a source of energy for the plasma acceleration in eruptive events (Low 1996; Forbes 2000). The most probable initial magnetic configuration of a CME is a flux rope consisting of twisted field lines (Chen, 1989; Lin et al. 1998; Titov & Demoulin 1999; Amari et al. 2000; Low 2001; Kliem & Török 2006; Zuccarello et al. 2012a). An alternative configuration of a CME source region is a sheared arcade (Moore & Roumeliotis 1992; Choe & Lee 1996; Antiochos et al., 1999), which is converted into a flux rope structure due to reconnection just before the eruption.

In this paper, we follow the idea that a flux rope exists in equilibrium in the coronal magnetic field for a rather long time before the eruption. van Tend & Kuperus (1978) showed first that there is a critical height for stable flux rope equilibria at which the background coronal magnetic field decreases faster than the inverse height. The transition from stability to instability was named later catastrophic loss of equilibrium and was assumed to be the cause of sudden eruptive events (Priest & Forbes 1990; Forbes & Isenberg 1991; Lin et al. 1998; Shmieder et al. 2013). van Tend & Kuperus (1978) modeled a flux rope with the magnetic field created by a straight line current. If a flux rope is curved, an additional force called the "Lorentz self-force" or "hoop force" is present (Bateman 1978). It is directed away from the curvature center. In the presence of an ambient magnetic field, the curved flux rope can be both in stable or unstable equilibrium depending on properties of the external field. Kliem & Török (2006) called the related instability "torus instability" and showed, following Bateman (1978), that it occurs when the background magnetic field decreases along the expanding flux rope major radius $R$ faster than $R^{-1.5}$. Demoulin & Aulanier (2010) carefully compared the two types of models and came to conclusion that the same physics is involved in the instabilities of circular and straight current channels. The stability of the flux-rope equilibrium in both models depends on the rate of the background field decrease, quantified by the so-called decay index,

$$n = -\frac{\partial \ln B_t}{\partial \ln h}, \qquad (1)$$

where $B_t$ is the horizontal magnetic-field component perpendicular to the flux rope axis and $h$ is the height above the photosphere. Demoulin & Aulanier (2010) found that for the typical range of current-channel thickness expected in the corona and used in many MHD simulations (Török & Kliem 2007; Schrijver et al. 2008; Fan 2010; Lugaz et al. 2011), and for a current channel expanding during an upward perturbation, a critical decay index $n_c$ has similar values for both the circular and straight current channels in the range 1.1 - 1.3.

Besides the cases of highly idealized models with translational and cylindrical symmetry, three-dimensional MHD simulations of the evolution of the magnetic field in the corona show that an eruption starts when the center of the flux rope reaches a critical height at which the corresponding potential field declines with height at a sufficiently steep rate, consistent with the onset of the torus instability of the flux rope (Fan 2010; Aulanier et al. 2010).

The coronal magnetic field is still largely elusive for reliable measurements (Tomczyk 2012). Photospheric magnetic field extrapolations are therefore commonly used for estimations of the value and structure of the coronal magnetic field. For the flux-rope stability analysis, information about the distribution of the external field (that is, the field produced by sources other than the flux rope current) is required. It is reasonable to assume that major coronal currents in the volume of interest are contained within the flux rope. For the magnetic field of currents below the photosphere, the coronal potential field is a rather good approximation. However, it should not be forgotten that the potential field does not represent the whole external field exactly.

Liu (2008) computed the field overlying the erupting filaments from the observed magnetic field at the Sun's surface based on a potential field source surface (PFSS) model (Schatten et al. 1969; Altschuler & Newkirk 1969; Hoeksema et al. 1982; Wang & Sheeley 1992). The decay index was computed after averaging the horizontal field along the polarity inversion line (PIL) or over the whole area of the active regions where the eruptions took place. In ten studied events, the decay index was in the range 1.1 - 1.7 for failed eruptions and in the range 1.4 - 2.2 for full eruptions. Liu et al. (2010) calculated spatial distribution of $n$ along the PIL using a nonlinear force-free field (NLFFF) extrapolation of an active region (Wheatland et al. 2000; Wiegelmann 2004). However, like in the work of Liu (2008), only a single value of $n$ was derived for the height range of 42–105 Mm. Along the not fully erupting section of the studied filament, the decay index was between 1.5 and 2.5.

Guo et al. (2010) obtained the height distribution of the decay index above the photosphere from a potential field extrapolation in the range 0.1 - 1.8 for an active region, which showed a confined eruption. They concluded that the decay index stayed below the torus instability threshold in the area where the erupting structure ascended. Cheng et al. (2011) through NLFFF extrapolation found that the decay index in the low corona (~ 10 Mm) is larger for eruptive flares than for confined ones. It generally increased from 0 at the surface to 2.5 at a height of ∼80 Mm. The decay index profile of the NLFFF model was significantly different from that of a potential field model only in the low corona. However, it is difficult to separate the background magnetic field, needed for the decay index calculation in the torus instability models, from the magnetic field induced by the current inside the flux rope in the NLFFF model. Kumar et al. (2012), based on a PFSS model, obtained the average decay index for the height range of 0.10 $R_{Sun}$ to 0.65 $R_{Sun}$ in active region NOAA 11163 as 1.74. They concluded that since the decay index exceeded the threshold of torus instability, the latter was the main driver of the CME. Nindos et al. (2012) determined from potential-field extrapolations as well as from the NLFFF extrapolations the decay index in active region NOAA 11158, which was the site of three major CMEs in February 2011. At the heights of CME onset, the decay indices were in the range 1.1 - 2.1. A NLFFF model of a sigmoidal region observed in 2007 February (Savcheva et al. 2012a, b) revealed that just before the occurrence of a B-class flare followed by a CME, the flux rope entered into the torus instability domain where the decay index of the potential arcade became 1.5. Xu et al. (2012) found that the mean decay index increased with CME speed for those CMEs with a speed below 1000 km s$^{-1}$ and stayed flat around 2.2 for the CMEs with higher speeds.

Filippov & Den (2001; 2002) calculated on the basis of photospheric magnetograms the decay index in the vicinity of filaments, using the potential magnetic field approximation. They defined a critical height $h_c$ as the height where $n = 1$ or, in other words, used $n_c$ for a straight current channel. They compared the measured heights of stable and eruptive filaments with the critical heights and found that the heights of stable filaments are usually well below the critical heights, while the heights of filaments just before their eruption are close to the instability threshold. This work had limited accuracy with respect to the simultaneous determination of the filament height and the critical height, because observations from a single point of view were used. The height of a filament can be easily measured when it is observed above the limb as a prominence. At that time the magnetic field in a photospheric area below the prominence cannot be obtained because the surface of the photosphere is nearly parallel to the light of sight. The magnetic field is measured most accurately when a region is located near the disk center. Using observations from a single point of view, one has to assume that the photospheric magnetic fields or the filament height do not change significantly during a period of a week. There are also some indirect methods to estimate the height of a filament above the photosphere when it is projected against the solar disk (d'Azambuja & d'Azambuja, 1948; Vrsnak et al., 1999; Zagnetko et al., 2005; Filippov & Zagnetko 2008). All of them have some specific limitations related to an assumption of small filament shape changes on a time scale of several days or to a supposed relationship between the filament shape and the potential magnetic field structure (Zagnetko et al. (2005) found that quiescent-filament plasma concentrated predominantly near the potential magnetic field neutral surfaces). After the launch of the *Solar Terrestrial Relations Observatory* (*STEREO*) mission, the 3D filament structure and the true process of filament eruption become accessible for study with unprecedented details (Li et al. 2010; Xu et al. 2010; Zapiór & Rudawy

2010; Gosain et al. 2009; Liewer et al. 2009; Bemporad 2011; Panasenco et al. 2011; Thompson 2011; Li et al. 2011; Sun et al. 2012).

In this paper, we analyze the eruptive event that occurred on 2010 October 21 near the central meridian of the Earth-side solar disk and was observed in detail by the *Solar Dynamic Observatory* (*SDO*) from the geosynchronous orbit as well as by both spacecraft of the *STEREO* mission separated by nearly ± 90º from the Earth along the Earth orbit. These simultaneous observations allow us both to measure accurately the height of the filament above the photosphere and to calculate the potential magnetic field in the corona.

## 2. OBSERVATIONS OF THE ERUPTION

We used data of the Atmospheric Imaging Assembly (AIA; Lemen et al. 2012) aboard the *SDO*, Sun Earth Connection Coronal and Heliospheric Investigation (SECCHI) EUVI (Wuelser et al. 2004; Howard et al. 2008) and coronagraphs COR 2 (Howard et al. 2008) aboard the *STEREO*, the Large Angle and Spectrometer Coronagraph (LASCO) C2 (Brueckner et al. 1995) and the Michelson Doppler Imager (MDI; Scherrer et al., 1995) aboard the *Solar and Heliospheric Observatory* (*SOHO*) as well as Hα observations of Kanzelhoehe Solar Observatory.

The long stable filament of the intermediate type was a very prominent feature in Hα images during its passage from the East limb to the central meridian on 2010 October 14-21. It was located on the eastern edge of AR 11113 and was tilted by an angle of about 30º with respect to the South-North direction (Figure 1). The AIA captured the start of the filament eruption around 13 UT on October 21 (Figure 2). Both *STEREO* spacecraft, Ahead and Behind, observed the prominence eruption above the Western and Eastern limb, respectively. Several 304 Å images obtained with SECCHI/EUVI are shown in Figure 2.

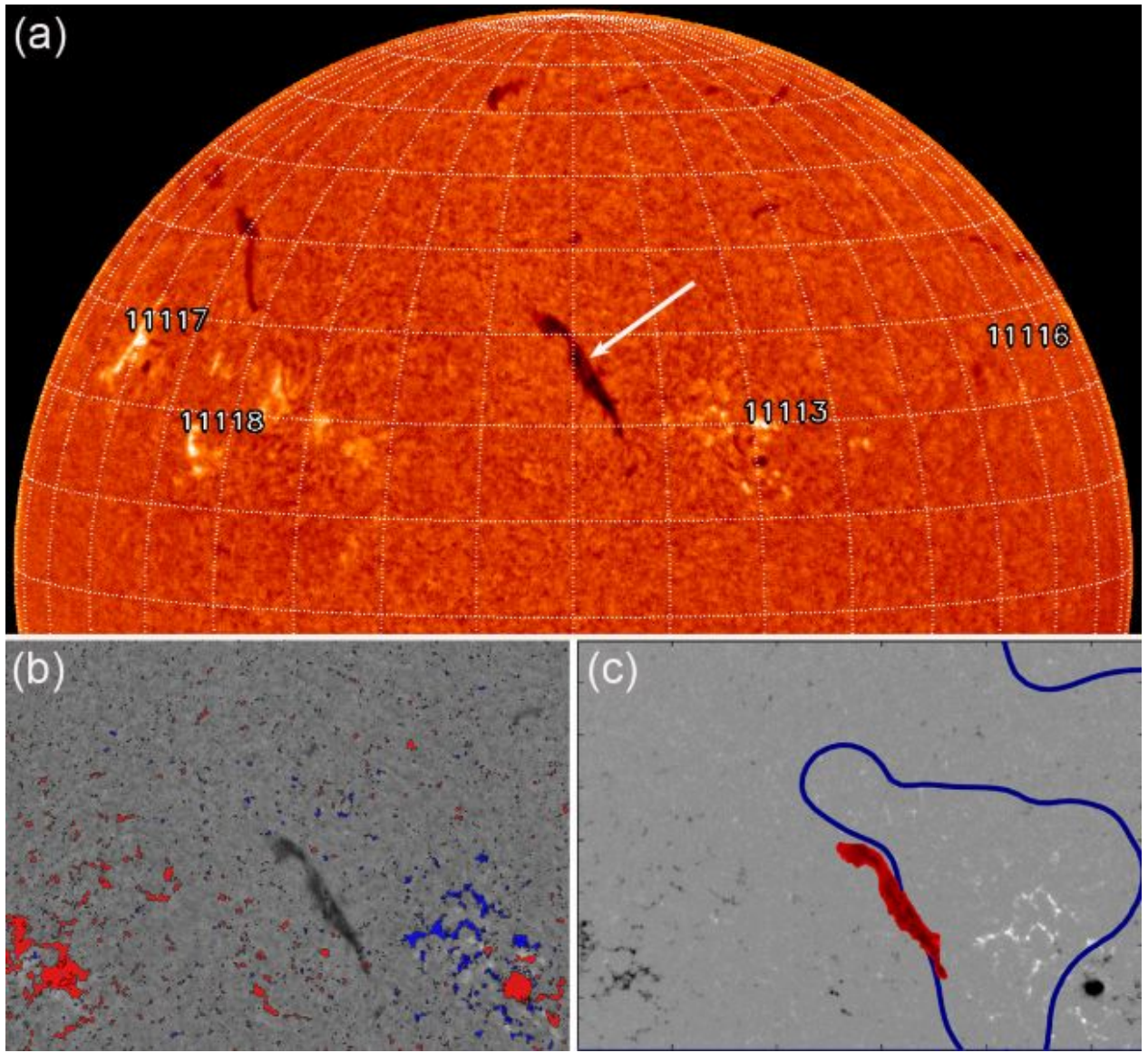

**Figure 1.** Position of the filament (white arrow) on the solar disk before the eruption (a). (Image from the site http://www.solarmonitor.org/). Position of the filament relative to the photospheric magnetic fields (b) and relative to the calculated polarity inversion line at the height of 48 Mm (c). The Hα filtergram in (a) is taken at 07:50 UT on October 21 at the Kanzelhoehe Solar Observatory. The *SOHO*/MDI magnetogram in (b) and (c) is taken at 08:02 UT. The size of the frame (c) is 880″ × 645″. (Courtesy of the Kanzelhoehe Solar Observatory and the *SOHO*/MDI consortium.)
(A color version of this figure and the following figures are available in the online journal.)

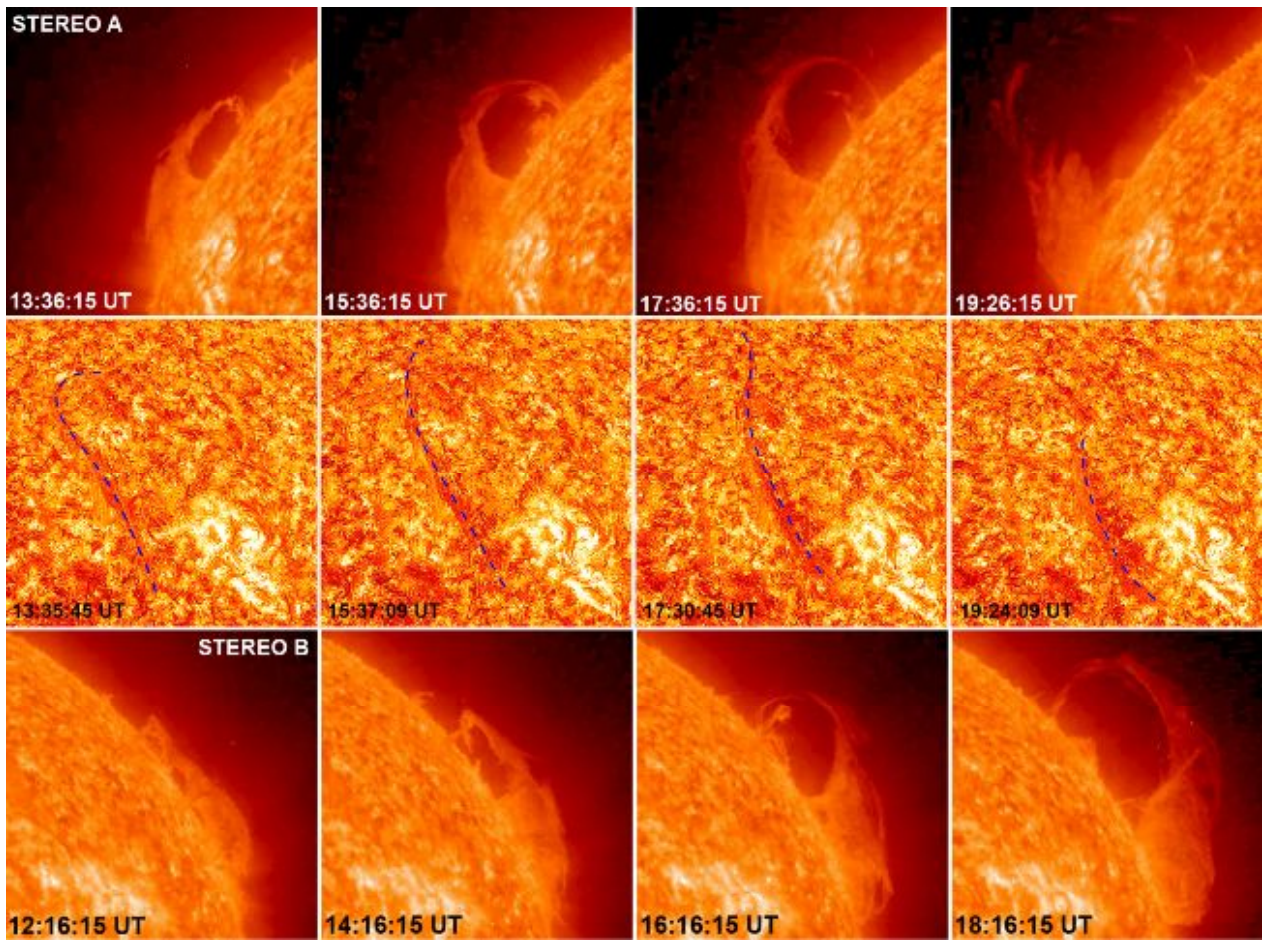

**Figure 2.** Observations of the eruptive filament on 2010 October 21 with *STEREO A*/SECCHI EUVI 304 Å (top row), with *SDO*/AIA 304 Å (middle row), and with *STEREO B*/SECCHI EUVI 304 Å. (bottom row). The thin dashed line indicates the filament spine. The size of each frame is 650" × 650". (Courtesy of the STEREO/SECCHI Consortium, NASA/SDO and the AIA science team.).

The associated CME appeared in the field-of-view of the space-born coronagraphs COR 2 and LASCO C2 around 00 UT on October 22 (Figure 3). The CME moved at an angle of about 45º to the equatorial plane and deviated to the West from the central-meridian plane as LASCO images show. According to the *SOHO*/LASCO CME Catalog[1], the CME appeared first in the field-of-view of C2 at 22:36 UT on October 21. It moved with a speed of 180 km s$^{-1}$ just above the occulting disk of C2 and reached a speed of 800 km s$^{-1}$ at the distance of 20 $R_{Sun}$. In the Catalog, the CME is characterized as a partial-halo-CME.

Figure 4 shows the height-time plot of the prominence top observed by both *STEREO* spacecraft. The height of the highest point of the prominence spine is measured above the limb. For *STEREO B* the filament initially was partly projected on the disk, therefore the height above the limb is lower than for *STEREO A*, which observed the prominence just above the limb. Since during the eruption the filament displaces to the West, later on it occupies a more symmetrical position relative to *STEREO A* and *B*, and the measured heights become equal. The height at which the filament starts to ascend rapidly is about 75 Mm. Uncertainties in the height measurements are mainly related to identification of the prominence top. The angular position of the highest point of the prominence spine was changing due to the changes in the prominence fine structure. Since the eruptive prominence changes its shape, tracing the top we do not follow the same plasma element. The large scattering in the values at 17-18 UT is caused by appearance and disappearance of individual prominence threads along its main body.

---

[1] http://cdaw.gsfc.nasa.gov/CME_list/

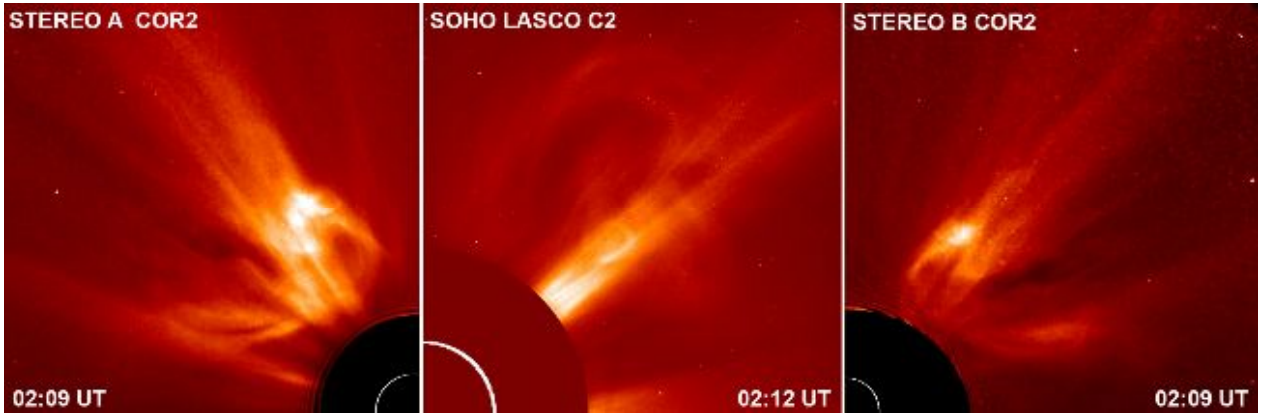

**Figure 3.** *STEREO*/COR2 and *SOHO*/LASCO C2 observations of the coronal mass ejection on 2010 October 22. Each frame is a quarter of the field-of-view of the corresponding coronagraph and has its own scale. The segments of white circles show the size of the solar disk. (Courtesy of the *STEREO*/SECCHI Consortium and the *SOHO*/LASCO Consortium, ESA and NASA).

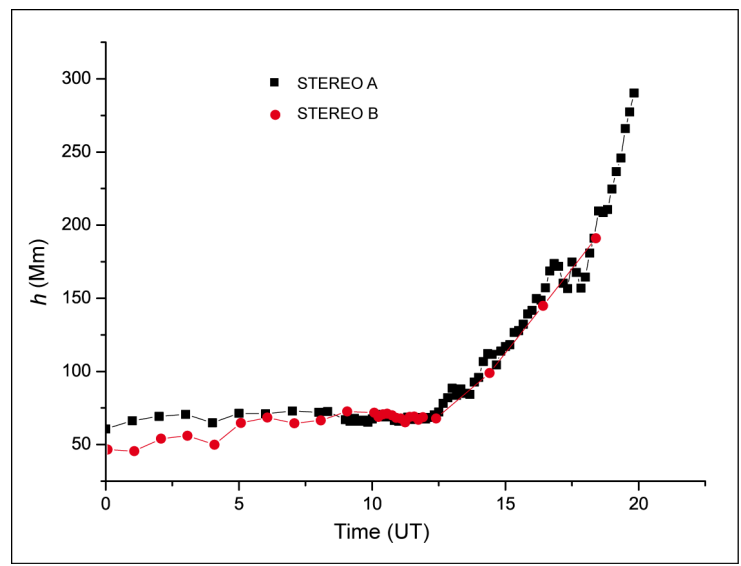

**Figure 4.** Height-time plot of the prominence top observed by the *STEREO* on 2010 October 21.

3. CORONAL MAGNETIC FIELD CALCULATIONS

The potential magnetic field in the corona can be found by solving the Neumann external boundary-value problem. There are a number of methods to solve numerically this problem (Altschuler & Newkirk 1969; Levine 1975; Adams & Pneuman 1976). The first code was developed and successfully used by Schmidt (1964). Since we are concerned with the magnetic field at the prominence height, which is small compared to the solar radius, we may neglect sphericity and use the well-known solution for half-space with a plane boundary (see, e.g., Tikhonov & Samarskii 1972, p. 363)

$$\mathbf{B}(x, y, z) = \frac{1}{2\pi} \iint_s \frac{B_n(x', y', 0)\, \mathbf{r}}{r^3}\, dx'dy', \qquad (2)$$

where $B_n(x',y',0)$ is the normal magnetic-field component on plane $S$, and $\mathbf{r}$ is the radius vector from a point on the surface to a given point in the corona. The $z$-axis is perpendicular to the boundary surface.

In numerical integration, the right-hand side of Equation (2) can be replaced by sums of small rectangular cells with the constant value of $B_n^{ij}$ within them and the coordinates of the cell boundaries $(x_i, y_j)$, $(x_{i+1}, y_{j+1})$, within which the integration can be performed analytically (Den et al. 1979):

$$B_x = \frac{1}{2\pi} \sum_{i,j} B_n^{ij} I_x^{ij}, \quad B_y = \frac{1}{2\pi} \sum_{i,j} B_n^{ij} I_y^{ij}, \quad B_z = \frac{1}{2\pi} \sum_{i,j} B_n^{ij} I_z^{ij}, \qquad (3)$$

where

$$I_x^{ij} = \ln \frac{(p_1 + \sqrt{m_2^2 + p_1^2 + z^2})(p_2 + \sqrt{m_1^2 + p_2^2 + z^2})}{(p_2 + \sqrt{m_2^2 + p_2^2 + z^2})(p_1 + \sqrt{m_1^2 + p_1^2 + z^2})},$$

$$I_y^{ij} = \ln \frac{(m_1 + \sqrt{p_2^2 + m_1^2 + z^2})(m_2 + \sqrt{p_1^2 + m_2^2 + z^2})}{(m_2 + \sqrt{p_2^2 + m_2^2 + z^2})(m_1 + \sqrt{p_1^2 + m_1^2 + z^2})},$$

$$I_z^{ij} = \mathrm{arctg}\, \frac{m_1 p_1}{z\sqrt{m_1^2 + p_1^2 + z^2}} + \mathrm{arctg}\, \frac{m_2 p_2}{z\sqrt{m_2^2 + p_2^2 + z^2}} -$$

$$- \mathrm{arctg}\, \frac{m_2 p_1}{z\sqrt{m_2^2 + p_1^2 + z^2}} - \mathrm{arctg}\, \frac{m_1 p_2}{z\sqrt{m_1^2 + p_2^2 + z^2}},$$

$$m_1 = x - x_{i+1}, \quad m_2 = x - x_i, \quad p_1 = y - y_{j+1}, \quad p_2 = y - y_j.$$

For the boundary condition, we used a full-disk magnetogram observed by SOHO/MDI. We cut out a rectangular area around the filament under study from the magnetogram of the full disk (Figure 5) and ignore the contribution of the magnetic sources in the areas outside it. This is justified if the region size is much larger than the filament height, or if the main sources of the field (active regions) lie within the cut-out region. Since the filament is located at a middle latitude (between N15 and N35), we should take into account the projection effect when we cut the rectangular area for the boundary condition for solving the Neumann external boundary-value problem.

If we cut a rectangular area at a middle latitude from a full-disk magnetogram, all pixels have different linear sizes on the photospheric surface. We construct the corrected array of equal-sized cells with equal angular latitudinal $\Delta\phi$ and longitudinal $\Delta\lambda$ dimensions. The field strength for each cell takes the value of the nearest pixel in the magnetogram with the corresponding latitude and longitude. To obtain the normal magnetic field component we multiply the measured line-of-sight field value by the cosine of the angle between the normal direction to the cell and the line-of-sight. In this way, we reconstruct the distribution of the magnetic field within the rectangular area as it would be observed if the region were at the center of the disk. Figure 1(c) shows an example of such a reconstructed magnetogram.

We calculate the distribution of the decay index using Equation (1) with $B_t = (B_x^2 + B_y^2)^{1/2}$ and Equations (3) at different heights. Thus we obtain the 3D distribution of $n$ within the domain surrounding the filament. It is well known that in a potential field, field lines are nearly perpendicular to a PIL. Thus, $B_t$ is a sufficiently good approximation of the field component perpendicular to the PIL, which is the one considered in the torus instability model. Only the contours $n = 1$ are shown in Figure 5. We expect that this is the critical value for filament stability based on our previous works. The areas where $n < 1$ are white, while the areas where $n > 1$ are yellow. Thick red lines represent PILs at the respective heights. The filament observed in Hα before the eruption occupies the space between the two short blues lines shown in the first panel of Figure 5 (compare with Figure 1). Obviously, the whole filament length at this height is within the area where $n < 1$. Zones of $n > 1$ increase in area at greater heights, because the magnetic field gradient on average increases with height.

We are interested in the distribution of the decay index along the section of the PIL occupied by the filament at the height at which the filament starts the rapid ascent. According to Figure 4, the filament starts to ascend rapidly at a height of about 75 Mm. The actual height of instability onset can be a little greater due to a projection effect, although the effect is not very significant due to the filament location close to the limb. From Figure 5 we can find that at a height of 80 Mm the contour of $n = 1$ touches the PIL which, according to the discussion presented below, should approximately coincide with the top part of the rising filament. This suggests that the filament becomes unstable when a part of it enters the region where $n > 1$. From this comparison, we can conclude that the critical decay index for the studied filament is about $n_c = 1$. The obtained value corresponds to the value found for a straight current channel and indicates that the effect of the flux-rope axis curvature (hoop force) is not significant in this event.

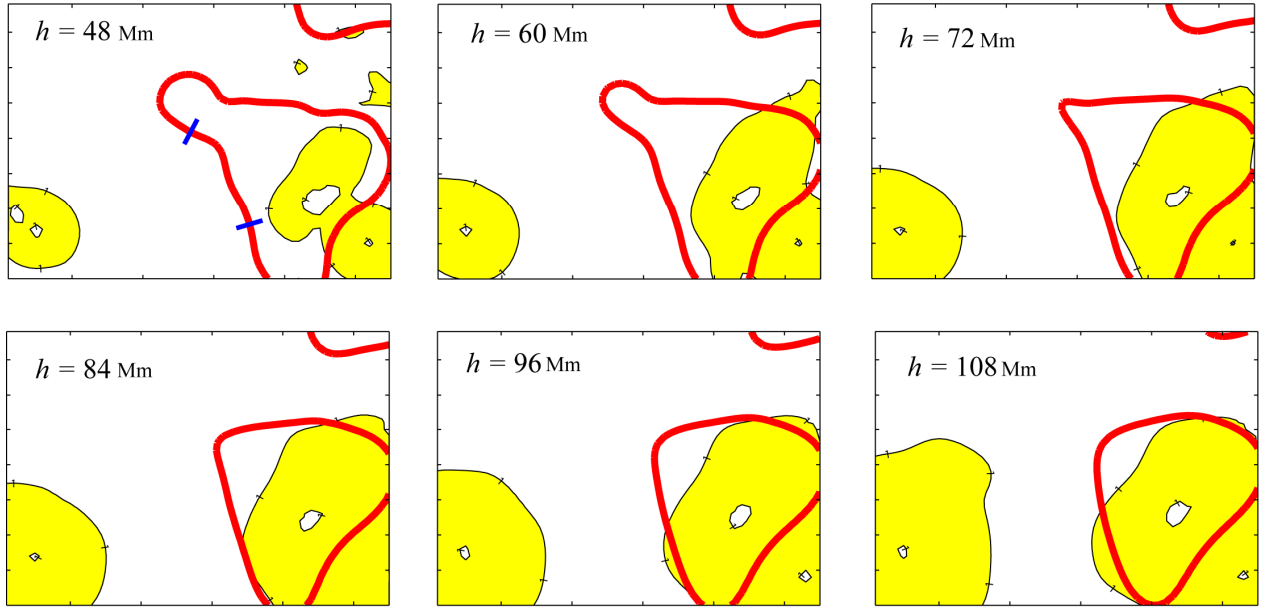

**Figure 5.** Distributions of the decay index and polarity inversion lines (thick red lines) at different heights. The size of each frame is 880" × 645".

Comparison of Figure 5 with the filament height-time plot in Figure 4 shows that the initial prominence height about 60 Mm is within the stability zone. The prominence height is slowly increasing in the early hours on October 21 with small oscillations. The start of the filament rapid ascending at the height of about 75 Mm corresponds to entering of the filament into the instability zone somewhere near 80 Mm.

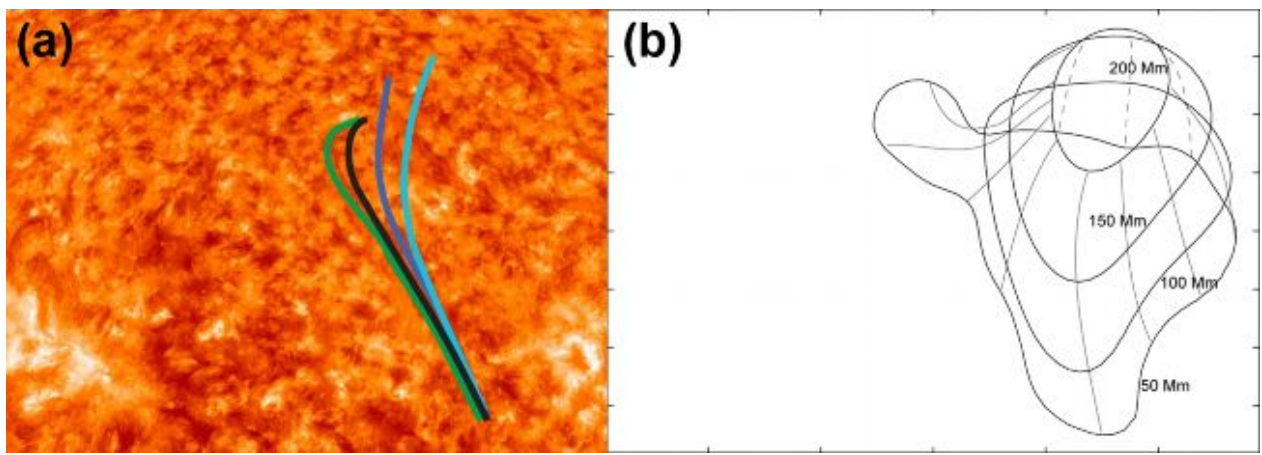

**Figure 6.** The filament spine at 13:08 UT (green line or the most-left line), 14:22 UT (black line), 17:10 UT (blue line), and 18:33 UT (cyan line or the most-right line) deduced from *SDO*/AIA 304 Å images (a). Calculated neutral surface (b).

After the loss of stable equilibrium, the flux rope moves under the action of the non-balanced Lorentz force and gravity. The vertical component of the Lorentz force is directed upward, while the horizontal component of the Lorentz force is directed toward the polarity inversion line (neutral line) at a given height. Therefore, until the inertia force influences significantly the eruptive filament motion, it should follow the magnetic neutral surface, the surface where $B_r = 0$. The shape of the calculated potential-magnetic-field neutral surface is shown in Figure 6(b). The declination of the perpendicular to the photospheric area from the line-of-sight by 30º is taken into account (compare with Figure 5 where the line-of-sight is assumed to be normal to the surface).

The observed positions of the filament spine at 13:08 UT, 14:22 UT, 17:10 UT, and 18:33 UT deduced from *SDO*/AIA 304 Å images are shown in Figure 6(a) (see also Figure 2). At these moments, the filament top was at the height of 80, 110, 160, and 210 Mm (Figure 4). The plot of the neutral surface in Figure 6(b) shows that the PIL is located more and more to the West as one goes up in height, and the same is true for the filament spine during its eruption, as can be seen in Figure 6(a). It suggests that on the initial stage the filament moves along the magnetic neutral surface, indicating that the shape of the neutral surface determines the direction of mass ejection. This is the reason why the CME propagates not along the central meridian plane but deviates to the West to be visible in the North-West quadrant of the LASCO field-of-view (Figure 3). Non-radial motion of flux ropes in latitudinal direction was simulated analytically by Filippov et al. (2001) for a solar eruptive filament. The deflection from the radial direction was due to the action of the coronal magnetic field on the net axial flux-rope electric current with the Lorenz force. On the early stage of the eruption, the flux rope moved nearly along the neutral surface. Flux-rope deflections were found in MHD numerical simulations (Aulanier et al. 2010; Lugaz et al. 2011; Török et al. 2011; Zuccarello et al. 2012b; Lynch & Edmondson 2013). In numerical models, sometimes it is not so easy to see what force causes the deflection. Shen et al. (2011) proposed that the deflection of a CME at an early stage may be caused by a nonuniform distribution of the background magnetic-field energy density and that the CME tended to propagate to the region with lower magnetic-energy density.

Sometimes, the magnetic axis of the flux rope in the CME rotates about the direction of ascent (Rust & LaBonte 2005; Green et al. 2007; Kliem et al. 2012). Usually it is considered as evidence for the MHD helical kink instability with a conversion of twist into writhe. Kliem et al. (2012) found in MHD simulations that the flux rope axis rotation was not guided by the changing orientation of the vertical field component's PIL with height. Possibly, this is a consequence of significant curvature of the flux rope in the vertical plane and rather strong external toroidal magnetic field in the model. The rotation is caused by the action of the horizontal component of the external toroidal magnetic field on the vertical sections of the axial flux-rope current (filament legs) as Isenberg & Forbes (2007) showed. Of course, a PIL does not represent the structure of the horizontal magnetic field.

## 4. CONCLUSIONS

We analyzed the filament eruption on 2010 October 21 observed from three different viewpoints (*STEREO A*, *B*, *SDO* and *SOHO*). The position of the filament at the beginning of the eruption was very favorable for the analysis of the filament stability on the basis of flux rope instability models. From the flux rope models described in the Introduction it follows that the flux rope becomes unstable when it arrives into ambient magnetic field with specific properties. The catastrophic loss of equilibrium (or torus instability) is believed to be the cause of filament eruptions. The filament was at the central meridian for an observer on the Earth not far from the center of the solar disk. This guarantees reliable photospheric magnetic field data within the area below the filament, obtained by *SOHO*/MDI, *SDO*/HMI, and ground-based solar magnetographs. Using MDI data we were able to calculate the potential magnetic field in the corona near the filament position and to obtain the distribution of the decay index of the magnetic field. For the *STEREO* spacecraft, the filament was seen above the limb as a prominence, which enables us to measure its height above the photosphere with high accuracy.

The prominence height was about 60 Mm on October 20 and was slowly increasing in the early hours on October 21 with small oscillations. The prominence started to ascend rapidly when it reached a height of about 75 Mm. The calculated distribution of the decay index indicates that the filament-carrying flux rope is in near-equilibrium when it is below 80 Mm and becomes unstable when its height exceed this value. So, these observations support the torus instability model and strongly indicate that the eruption begins when the flux rope approaches the stability threshold. Our results indicate that the critical decay index was close to one for the eruption studied here.

The changes in shape and position of the eruptive filament provide evidence that it moved along the magnetic neutral surface during the initial stage of the eruption. Therefore, the structure of the magnetic field around a filament determines the direction of mass ejection. This is significant for estimation of geoeffectiveness of filament eruptions and associated CMEs.

The author thanks the referee for helpful comments and suggestions that substantially improved the paper. The author thanks the Kanzelhoehe Solar Observatory, *STEREO*, *SOHO*, and *SDO* teams for the high-quality data supplied. This work was supported in part by the Russian Foundation for Basic Research (grants 12-02-00008 and 12-02-92692) and the Program # 22 of the Russian Academy of Sciences.